# Coded-Aperture Imaging Using Photo-induced Reconfigurable Aperture Arrays for Mapping Terahertz Beams

Akash Kannegulla[1], Zhenguo Jiang[1], Syed Rahman[1], Patrick Fay[1], Huili Grace Xing[1], Li-Jing Cheng[2*], Lei Liu[1*]

[1]Department of Electrical Engineering, University of Notre Dame, Notre Dame, IN 46556, USA. Email: lliu3@nd.edu

[2]School of Electrical Engineering and Computer Science, Oregon State University, OR 97311, USA. Email: chengli@eecs.oregonstate.edu

**Abstract** – We report terahertz coded-aperture imaging using photo-induced reconfigurable aperture arrays on a silicon wafer. The coded aperture was implemented using programmable illumination from a commercially available digital light processing projector. At 590 GHz, each of the array element apertures can be optically turned on and off with a modulation depth of 20 dB and a modulation rate of ~1.3 kHz. Prototype demonstrations of 4 × 4 coded-aperture imaging using Hadamard coding have been performed and this technique has been successfully applied to mapping THz beams by using a 6 × 6 aperture array at 590 GHz. The imaging results agree closely with theoretical calculations based on Gaussian beam transformation, demonstrating that this technique is promising for realizing real-time and low-cost terahertz cameras for many applications. The reported approach provides a simple but powerful means to visualize THz beams, which is highly desired in quasi-optical system alignment, quantum-cascade laser design and characterization, and THz antenna characterization.

The submillimeter-wave and terahertz (THz) region in the electromagnetic spectrum has become more and more important to radio astronomy, chemical spectroscopy, bio-sensing, medical imaging, security screening and defense [1-4]. In recent years, technologists have intensified their efforts to develop imaging systems operating in the THz region for all the above applications [5-8]. In addition, high-performance and low-cost imaging devices are highly desired for THz beam visualization for scientific metrology applications such as quasi-optical system alignment, quantum-cascade laser (QCL) design and optimization [9], and THz antenna characterization [10]. To date, the THz imaging systems that have been demonstrated generally fall into one of three categories: (1) single-element imagers that obtain images by mechanical scanning, (2) array imagers (e.g. focal-plane arrays (FPAs)) that consist of an array of imaging sensor elements [6-8], and (3) coded-aperture imaging (CAI) using two-dimensional aperture masks [11-14]. In many applications, important events happen on the scale of microseconds, making imaging by mechanical scanning impractical due to the inherently low frame rates. Array-based imagers such as FPAs can greatly reduce observing and processing time by recording imaging information in parallel. Although FPAs offer the highest imaging speed and signal-to-noise ratio, they tend to be complicated and expensive, especially for large-scale arrays with high imaging resolution.

Compared to THz imaging using mechanical scanning and focal-plane arrays, CAI offers the advantage of both high performance (i.e., high signal-to-noise ratio and high frame rate) as well as the potential for realizing simple and low-cost systems. CAI-based systems are based on spatial encoding and modulation to eliminate the need for detector arrays. In this imaging technique, a single THz detector in combination with a series of $N$ x $N$ coded aperture masks is required in order to obtain an image with an $N$ x $N$ resolution. Measurements with $N^2$ masks are taken, and the same number of linear equations are then solved to reconstruct the object image [11]. This basic concept has been demonstrated using masks fabricated on printed circuit boards [12]. In order to achieve high frame rates with THz CAI, aperture arrays electronically actuated by Schottky diodes [13] and graphene modulators [14], respectively, have been proposed to realize the required coded masking. However, these approaches require complicated and prepatterned circuits for operation, which also results in expensive and complex systems.

In this letter, we report THz CAI using photo-induced reconfigurable aperture arrays using an unpatterned silicon wafer illuminated by a commercially available digital light processing projector (DLP). The optical THz modulation mechanism described in [15] was employed to spatially modulate each array pixel. Pixels illuminated with light will be turned "off" due to increased photo-induced free carriers and local conductivity, while other pixels remain highly transparent ("on") to THz signal. This approach allows us to generate extremely large-scale reconfigurable coded masks (e.g., 1024 × 768, only limited by the DLP resolution and carrier diffusion lengths in the Si wafer) for THz CAI, without the need for any microfabrication processes or precise alignments. We demonstrate that at 590 GHz, each array element aperture (pixel) can be effectively turned "on" and "off" with a modulation depth of 20 dB and at a modulation rate of ~1.3 kHz. Prototype demonstrations of 4 × 4 coded-aperture imaging using Hadamard coding [11] have been performed and this technique has been successfully applied to mapping THz beams with 6 × 6 pixels at 590 GHz.

Fig. 1 shows the experimental setup [15, 16] in which a frequency multiplication chain based on Schottky diode multipliers (Virginia Diodes, Inc.) was used as a THz source in the frequency range of 570-600 GHz with an average output power of approximately 1mW. The output was coupled to free space using a WR-1.5 horn antenna. Four off-axis parabolic mirrors (M1-M4 in Fig. 1) were used to collimate and focus the THz beam. After passing through the Si wafer that serves as the coded aperture, the focused THz beam was reflected by an indium tin oxide (ITO) coated glass plate onto a broadband zero-bias Schottky diode

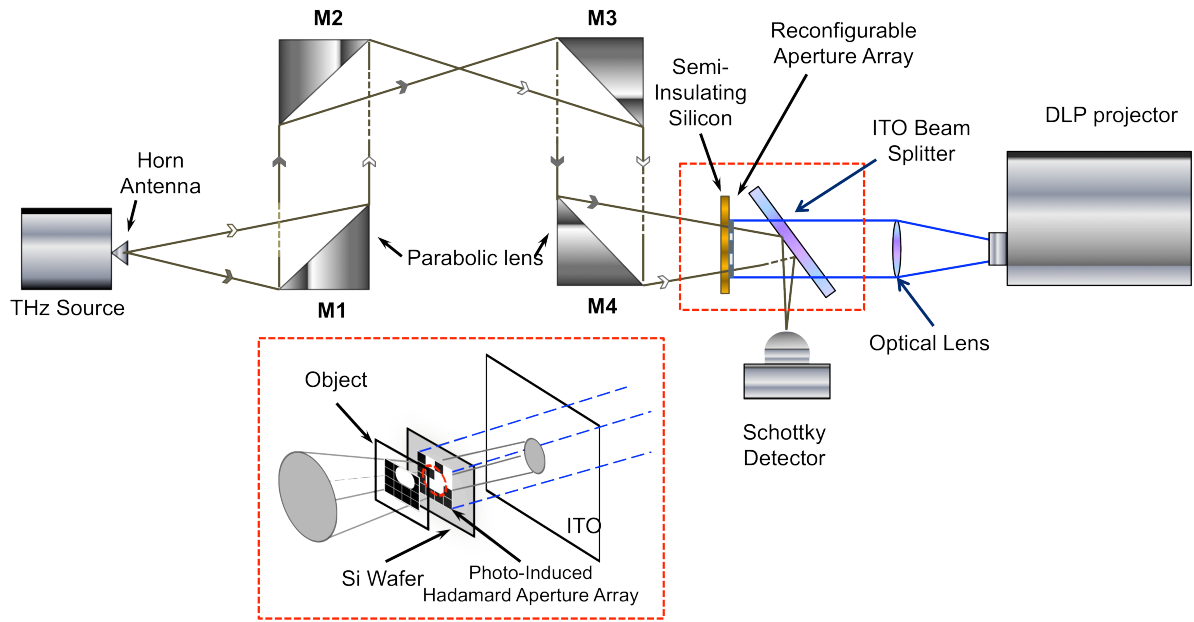

Fig.1. (color online) Diagram of the experiment setup. Reconfigurable photo-induced Hadamard aperture array was produced on a semi-insulating silicon wafer for THz coded-aperture imaging at 590 GHz. A commercially available DLP project was employed for generating the Hadamard patterns.

(ZBD) detector [17]. The ITO-coated plate was mounted at 45° with respect to the THz beam; the ITO-coated plate reflects the THz signal, but is optically transparent so that the optical pattern from the DLP can project onto the Si wafer. A 200 μm thick double-side polished semi-insulating silicon wafer was used as the coded-aperture modulator. This silicon wafer was illuminated by a commercially available DLP to generate reconfigurable Hadamard coded aperture arrays for CAI (see Fig. 1 inset for a representative 4 × 4 Hadamard array). The DLP system consisted of a 0.7" digital mirror device (DMD) panel with 1024 × 768 dot resolution and a 200 W mercury lamp. For this application, the RGB filter wheel was removed to allow generating white light images with a maximum brightness of ~2500 lumens after passing through a visible bandpass filter (400-800 nm). To focus the projected image down to an area of 10 × 10 mm$^2$ on the Si wafer, an additional lens was inserted in the optical path. For the imaging demonstration, a thin metal target (covered by absorber) with an open aperture was placed in the THz beam just before the silicon wafer (position "O" in the inset of Fig. 1) to serve as an object. For the THz beam mapping application, the silicon wafer was moved axially along the THz beam while keeping the aperture array patterns focused (the DLP moved together with the silicon wafer).

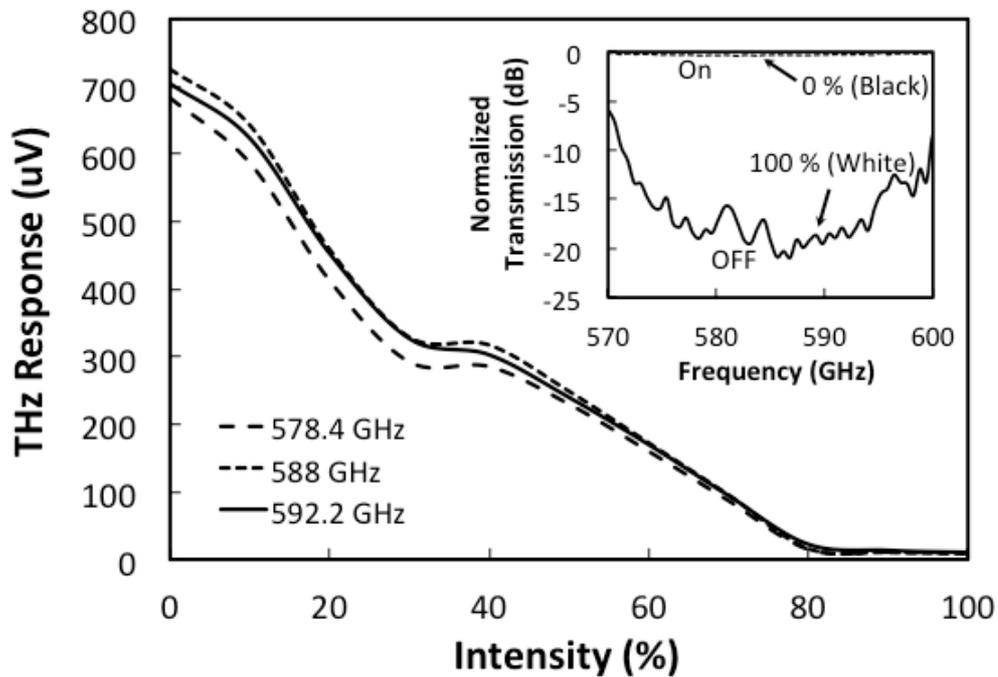

Fig. 2: (a) Measured THz responses from the detector as a function of DLP light intensity at three frequencies, i.e., 578.4, 588 and 592.2 GHz. The inset shows the normalized transmission for 0% and 100% intensity measured over the entire frequency range of 570-600 GHz. A modulation depth, or "on" and "off" ratio of ~ 20 dB has been obtained [15].

Previous demonstrations have shown that free-carrier absorption in Si can be effective for optically-controlled modulation of continuous waves in the frequency range of 570-600 GHz [15]. In this work, we have extended this concept to spatially encode the array pixels for reconfigurable Hadamard aperture masks. Fig. 2 shows the measured THz responses from the detector as a function of DLP light intensity (0%-100%, 100% corresponding to white light at ~2500 lumens) for three THz frequencies (i.e., 578.4 GHz, 588 GHz and 592.2 GHz). With increasing of the photo-excitation intensity, the density of free carriers increases, resulting in reduced transmitted THz power (power is proportional to the detector output voltage in the square-law region of the detector). The inset of Fig. 2 shows the normalized THz transmission (normal to the response without light) for 0% and 100% intensity measured over the entire frequency range of 570-600 GHz. A modulation depth, or the "on" (with 0% or black light) and "off" (with 100% or white light) ratio of ~ 20 dB at 585 GHz has been obtained. A relatively flat modulation depth is obtained over the frequency range of 578-592 GHz.

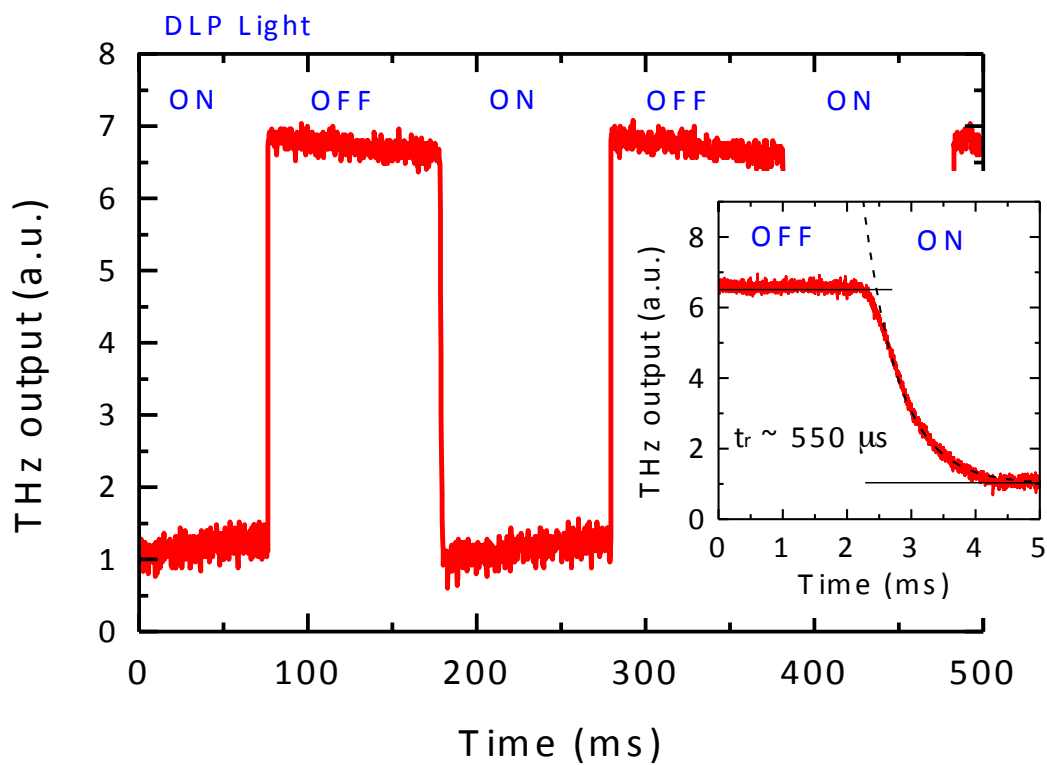

Fig. 3: Time response of THz wave modulation by switching DLP projection light at a rate of 5 Hz. The inset shows the enlarged view of transition. The transition time was determined by the switching rate of DMD panel and was measured to be ~550 μs.

To estimate the modulation speed, we illuminated uniform white light (100%) pulses onto the silicon wafer by programming the DLP to "flash" at a frequency of 5 Hz. Figure 3 shows the modulated THz output signals (585 GHz) corresponding to the photo-excitation from the

DLP. The transient response (Fig. 3 inset) of the temporal THz modulation shows a 10-90% rise (from "off" to "on") time of ~550 μs, corresponding to a 3-dB bandwidth of ~1.3 kHz. This modulation speed is slower than what is expected based on the free carrier recombination rate in silicon [18], but rather could be limited by the speed of the DMD array control electronics (~1.9 kHz) in the DLP projector [19]. High speed DMD chipsets such as DLPC410 by Texas Instruments, Inc. could be employed to provide a 32 kHz frame rate for much improved imaging speed [19].

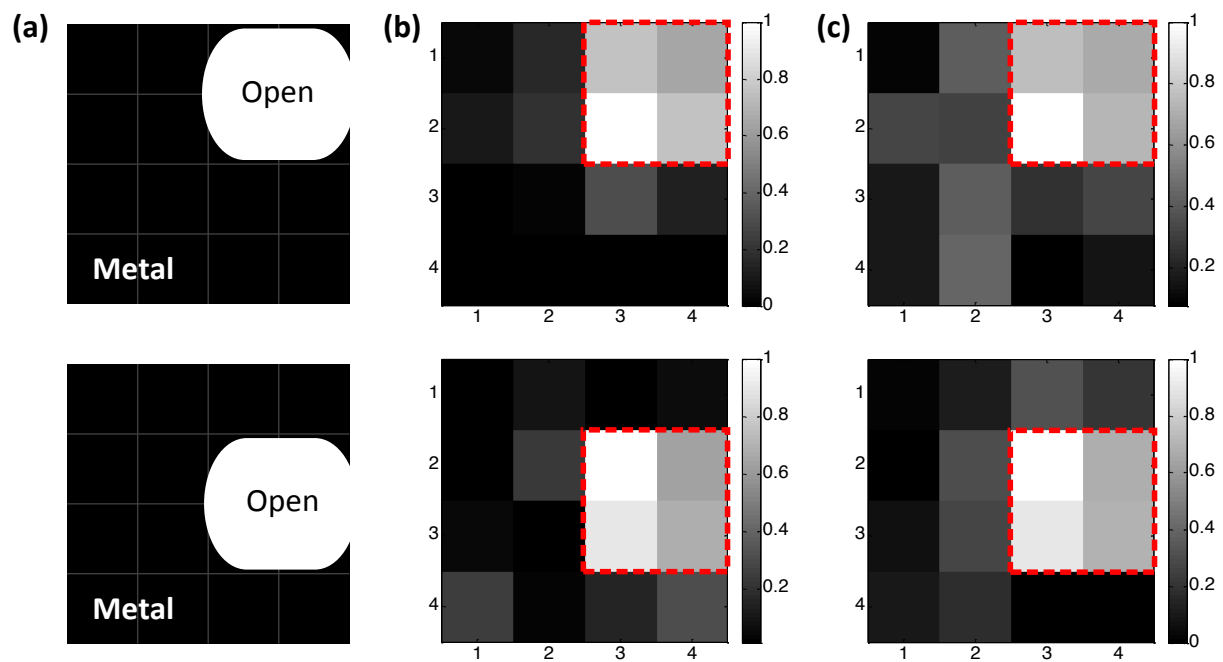

Fig. 4: 4 x 4 THz CAI at 590 GHz using reconfigurable photo-induced aperture arrays on a silicon wafer. (a) Left column: thin metal pieces (covered by absorber) with aperture openings at different positions as imaging objects. (b) Middle column: Single aperture CAI results. (c) Right column: Hadamard CAI imaging results.

For a prototype demonstration of THz CAI using the photo-induced reconfigurable aperture arrays (as seen in Fig. 1), we performed THz imaging at 590 GHz with a 4 x 4 pixel array. In this experiment, we employed the well-known Hadamard coding [11] (4 x 4 Hadamard matrix containing "1" and "-1" elements) for generating a series of coded masks. In order to generate the "-1" elements in the Hadamard matrix, we recorded the result corresponding to each matrix using an array with elements "1" (aperture "on") and "0" (aperture "off") and its complementary array, and then subtracted the two measurements from each other. In this way, a total number of 32 arrays were used for imaging with 4 x 4 pixels. The use of compressed CAI [12] can significantly reduce the required number of masks. To verify the effectiveness of this imaging approach, a thin metal target (covered by absorber) with an open aperture (see Fig. 1 inset) was used as the test object.

The imaging was performed automatically with a LabView program. Each array pattern was illuminated onto the silicon wafer for one second followed by a full black image for another second to reduce the heat effect on the silicon wafer introduced by the DLP projector. The above procedure was repeated with reconfigured Hadamard patterns for 32 times. 16 linear equations were then built and solved to reconstruct the image for the object. For comparison, imaging results using the single aperture CAI as described in [15] are shown in Fig. 4(b) and shown in Fig. 4(c) are the imaging results using the Hadamard coding for two different objects. The reconstructed images agree quite well with the original objects showing this technique is promising for THz imaging. Since the modulation speed of this approach was estimated to be ~1.3 kHz, a real-time video-rate (30 frame/sec) imaging with 7 x 7 resolution can be potentially demonstrated. A real-time low-cost THz camera for many practical applications could be realized on the basis of this approach.

In addition, we applied the above THz imaging approach to map the THz beam in the experimental setup as shown in Fig. 1. The THz beam waist (or radius $w$) size as a function of distance from the parabolic mirror M4 ($d$) was first calculated using Gaussian beam transformation described in [20] and the result is shown in Fig. 5 (a). For mapping the THz beam, the imaging object (thin metal piece) for the prototype CAI demonstration was removed from the system. The silicon wafer was moved to three different positions P1 ($d_1$ = 88 mm), P2 ($d_2$ = 92 mm) and P3 ($d_3$ = 96 mm) as shown in the inset of Fig. 5(a). During this process, the projected image from the DLP was kept focused (DLP moved together with the silicon wafer) and the imaging area on the silicon wafer remained 10 x 10 mm$^2$. As shown in Fig. 5(b), THz images at the three positions were taken using the above approach with 6 x 6 pixels. Although with relatively low resolution, all three images clearly show the THz beam at the positions where the silicon wafer was placed with a brighter region at the imaging area center. The red dashed circles (~4.5 mm, 3.5 mm and 2.5 mm radius for P1, P2 and P3 respectively) show the calculated THz beam sizes based on the calculation in Fig. 5(a), indicating that good agreement between the theoretical calculation and the experiment has been obtained. This suggests that this approach is a simple but potentially powerful means to visualize THz beams in a quasi-optical system. This same technique could be quickly refined and applied to quantum-cascade laser optimization and characterization, as well as THz antenna characterization. Future work will be focused on the development of real-time and low-cost THz cameras with more sophisticated coding, specifically designed DLP chips/systems, and high-performance THz detectors and receivers [6, 7].

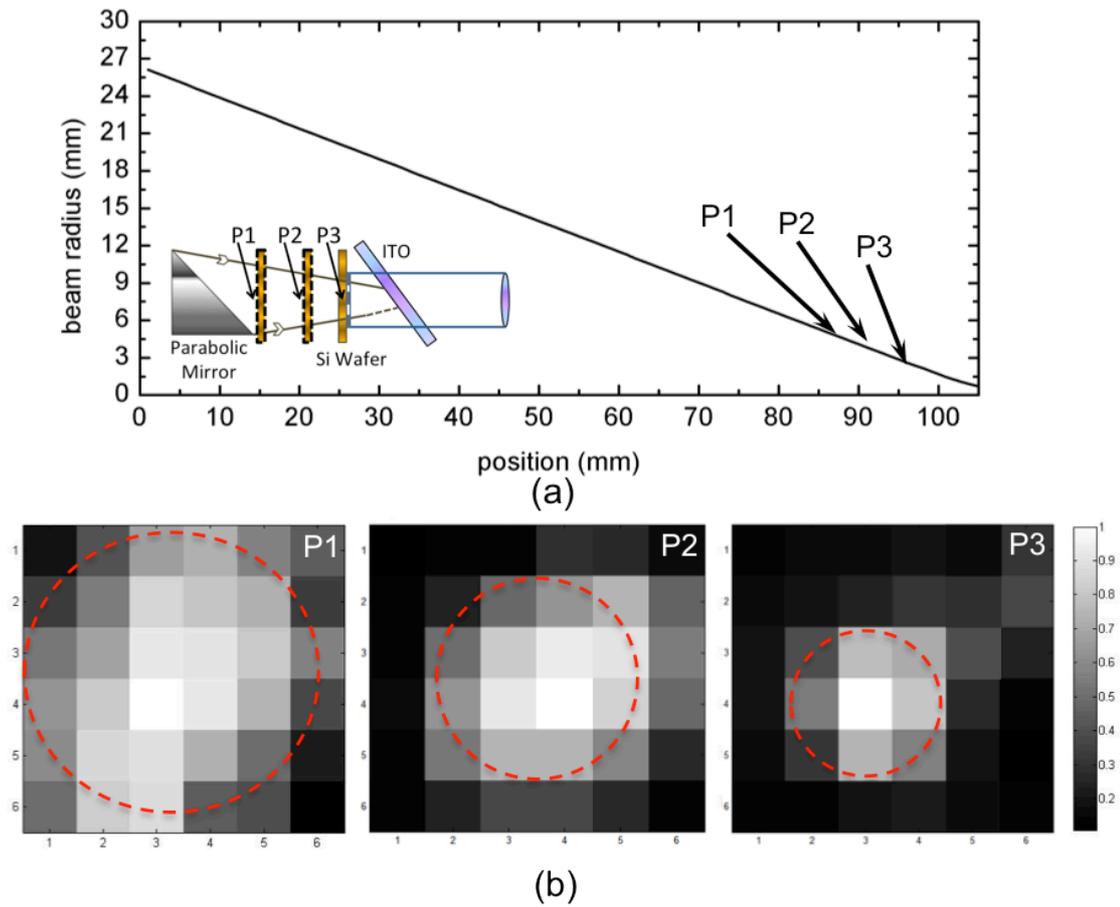

Fig. 5. (a) Calculated THz beam waist as a function of the distance from the parabolic mirror M4. The inset shows the three positions chosen for taking THz images using the reported THz CAI approach. (b) Images of the THz beam in the quasi-optical system (shown in Fig. 1) at the three positions indicated in (a). Red dashed circles show the calculated THz beam sizes (~ 9 mm, 7 mm, and 5 mm for P1, P2, and P3, respectively) using Gaussian beam transformation [19].

In conclusion, we report a novel approach for terahertz CAI using photo-induced reconfigurable aperture arrays on a silicon wafer illuminated by a commercially available DLP projector. At 590 GHz, each of the array element aperture can be optically turned on and off with a modulation depth of 20 dB and a speed of ~1.3 kHz. Prototype demonstrations of 4 × 4 coded-aperture imaging using the Hadamard coding have been performed and this technique has been successfully applied to mapping THz beams with 6 × 6 pixels at 590 GHz. The reported approach provides a simple but powerful means to visualize THz beams, which is highly desired in quasi-optical system alignment, quantum-cascade laser design and THz antenna characterization.

**Acknowledgement**

This work was partially supported by NSF Grants ECCS-1002088, ECCS-1102214 and ECCS-1202452. The authors also would like to acknowledge partial supports from the Advanced Diagnostics and Therapeutics (AD&T) and the Center for Nano Science and Technology (NDnano) at the University of Notre Dame.

**Reference:**

[1] B. Hu, M. Nuss, Optics Letters **20**, 1761 (1995).

[2] T. G. Phillips, J. Keene, Proceedings of the IEEE **80**, 1662-1678 (1992).

[3] A. Markelz, A. Roitberg, E. J. Heilweil, Chemical Physics Letters **320**, 42-48 (2000).

[4] E. Brown, D. Woolard, A. Samuels, T. Globus and B. Gelmont, IEEE MTT-S Int. Microwave Symp. Digest **3**, 1591-1594 (2002).

[5] D. B. Rutledge, and M. S. Muha, IEEE Trans. Antennas Propagat. **30**, 535-540 (1982).

[6] L. Liu, H. Xu, A. W. Lichtenberger, and R. M. Weikle, II, IEEE Trans. Microwave Theory Tech. **58**, 1943-1951 (2010).

[7] S. Rahman, Z. Jiang, Y. Xie, H. Xing, P. Fay, L. Liu, 23th Int. Symp. Space THz Tech., Tokyo, Japan (2012).

[8] D. J. Burdette, J. Alverbro, Z. Zhang, P. Fay, Y. Ni, P. Potet, K. Sertel, G. Trichopoulos, K. Topalli, J. Volakis, H. Lee Mosbacker, SPIE proceedings **8023**, (2011).

[9] M. Cui, J. N. Hovenier, Y. Ren, N. Vercruyssen, J. R. Gao et al., Appl. Phys. Lett. **102**, 111113 (2013).

[10] C. H. Smith, III, H. Xu, J. L. Hesler, N. S. Barker, 33rd International Conference on Infrared, Millimeter and Terahertz Waves (IRMMW-THz), Pasadena, CA (2008).

[11] I. Valova, Y. Kosugi, IEEE Trans. Information Technology in Biomedicine **4**, 306-19 (2000).

[12] W. L. Chan, K. Charan, D. Takhar, K. F. Kelly, R. G. Baraniuk, and D. M. Mittleman, Appl. Phys. Lett. **93**, 121105 (2008).

[13] S. Hawasli, N. Alijabarri, R. M. Weikle II, 37[th] International Conference on Infrared, Millimeter, and Terahertz Waves, Wollongong, NSW, Australia, (2012).

[14] B. Sensale-Rodriguez, S. Rafique, R. Yan, M. Zhu, V. Protasenko, D. Jena, L. Liu, H. G. Xing, Opt. Express **21**, 2324-2330 (2013).

[15] L. Cheng, L. Liu, Opt. Express, submitted (2013).

[16] L. Liu, R. Pathak, L.-J. Cheng, and T. Wang, Sensors and Actuators B **184**, 228-234 (2013).

[17] L. Liu, J. L. Hesler, H. Xu, A. W. Lichtenberger, R. M. Weikle, II, IEEE Microwave and Wireless Components Letters **20**, 504-506 (2010).

[18] H. Alius, G. Dodel, Infrared Phys. **32**, 1-11 (1991).

[19] http://www.ti.com/lit/ds/symlink/dlpc410.pdf

[20] P. F. Goldsmith, "Quasioptical systems: Gaussian beam quasi-optical propagation and applications," Wiley & Sons, Inc. (1997).